\documentclass{aa}  

\usepackage{graphicx}
\usepackage{txfonts}
\usepackage[breaklinks=true]{hyperref}
\usepackage{natbib} 
\bibpunct{(}{)}{;}{a}{}{,} 

\usepackage{verbatim}
\usepackage{url}
\usepackage{epstopdf}

\begin{document}

   \title{On the age of Galactic bulge
     microlensed dwarf and subgiant stars}

   \subtitle{}

   \author{G. Valle \inst{1,2,3}, M. Dell'Omodarme \inst{3}, P.G. Prada Moroni
     \inst{2,3}, S. Degl'Innocenti \inst{2,3} 
          }

   \authorrunning{Valle, G. et al.}

   \institute{
INAF - Osservatorio Astronomico di Collurania, Via Maggini, I-64100, Teramo, Italy 
\and
 INFN,
 Sezione di Pisa, Largo Pontecorvo 3, I-56127, Pisa, Italy
\and
Dipartimento di Fisica ``Enrico Fermi'',
Universit\`a di Pisa, Largo Pontecorvo 3, I-56127, Pisa, Italy
 }

   \offprints{G. Valle, valle@df.unipi.it}

   \date{Received 05/02/2015 ; accepted 03/03/2015}

  \abstract
{Recent results by Bensby and collaborators on the ages of microlensed dwarf and subgiant stars in the Galactic bulge have challenged the picture of an exclusively old stellar population, because  ages significantly younger than 9 Gyr have been found.} 
   {  
However,  these age estimates have not been independently confirmed with different techniques and theoretical stellar models. One of the aims of this paper is to verify these results by means of a grid-based method. We also quantify the systematic 
biases that might be induced by some assumptions adopted to compute stellar models. In particular, we explore the impact of increasing the initial helium abundance, neglecting the element microscopic diffusion, and changing the mixing-length calibration in theoretical stellar track computations.
}
{  
We adopt the SCEPtER pipeline with a newly computed stellar model grid for metallicities [Fe/H] from $-2.00$ dex to $0.55$ dex, and masses in the 
range [0.60; 1.60] $M_{\sun}$ from the ZAMS to the helium flash at the red giant branch tip. By means of Monte Carlo simulations we show  for the considered evolutionary phases that our technique provides unbiased age estimates.}
  {
Our age results are in good agreement with Bensby and collaborators findings and show 16  stars younger than 5 Gyr and 28 younger than 9 Gyr over a sample of 58. The effect of a helium enhancement as large as $\Delta Y/\Delta Z = 5$ 
is quite modest, resulting in a mean age increase of metal rich stars of 0.6 Gyr. Even simultaneously adopting a high helium content and the upper values of age estimates, there is evidence of 4 stars younger than 5 Gyr and 15 younger than 9 Gyr. 
For stars younger than 5 Gyr, the use of stellar models computed by neglecting microscopic diffusion or by assuming a super-solar mixing-length value leads to a mean increase in the age estimates of about 0.4 Gyr and 0.5 Gyr respectively. 
Even considering the upper values for the age estimates, there are four stars estimated younger than 5 Gyr is in both cases. Thus,
the assessment of a sizeable fraction of  young stars among the microlensed sample in the Galactic bulge appears robust. 
 }
{}

   \keywords{
Stars: evolution --
gravitational lensing: micro --
Galaxy: bulge --
methods: statistical -- 
Galaxy: evolution
}

   \maketitle

\section{Introduction}\label{sec:intro}

Recently a debate on the age of the stars in the Galactic bulge has  developed. 
Until few years ago, the picture of a Galactic bulge composed of old stars was commonly accepted. The evidence  from Hubble Space Telescope photometry of the Galactic bulge main-sequence turnoff points toward an exclusively old stellar population, with ages greater than about 10 Gyr 
\citep[e.g.][]{Zoccali2003,Brown2010,Clarkson2011, Valenti2013}. The similarity between the Galactic bulge and Galactic globular clusters suggested an age estimate of 12-13 Gyr \citep{Marin2009}.

However, in recent years a different scenario coming from gravitational microlensing results was proposed. \citet{Bensby2011,Bensby2013} obtained high-resolution spectra of 58 bulge main-sequence and subgiant stars, from which they derived measurements of $T_{\rm eff}$, $\log g$, and [Fe/H]. Surprisingly, the comparison of the observations with theoretical isochrones 
 revealed a significant fraction of young and intermediate-age stars. In fact, the \citet{Bensby2013} results showed the presence of 13 stars out of 58 (22\%) younger than 5 Gyr and 31 stars (53\%) younger than 9 Gyr.
 
As a consequence of these results, there has been some effort in the literature     
 to reconcile the two pictures.
A  first attempt was made by \citet{Nataf2012}, who suggested that taking into account the
possible Galactic bulge helium enhancement, the differences between the
two sets of age estimates should vanish. However, this result was questioned by
\citet{Bensby2013}, who claimed that the effect of the helium enhancement is
insufficient to contradict the conclusion of a substantial presence of stars
at young and intermediate ages. Furthermore, a possible contamination of the
microlensing sample by disk star and bias toward a lower age was proposed
\citep{Nataf2013}. However, none of these distortions seem to be large enough
to refute the \citet{Bensby2013} findings. More recently, \citet{Ness2014}
suggested that the discrepancy could be due to the different spatial
distributions in 
the Galactic bulge of old and young stars.

Given the importance of the age estimates by \citet{Bensby2013} in the framework of Galactic bulge formation scenarios \citep[see e.g.][and references therein]{Ness2014}, we feel that an independent analysis of these results -- adopting a different estimation technique and different theoretical stellar models -- and an analysis of potential systematic biases is mandatory.
The purpose of the present work is to provide such an independent check  adopting the SCEPtER pipeline \citep{scepter1, eta}. 

We also explore in detail the impact on these age estimates of changing some parameters or prescriptions required to compute a grid of stellar models that are not yet tightly constrained. In particular, we analysed the effect 
of a helium-rich Galactic bulge by computing a grid of stellar models with a helium-to-metal enrichment ratio $\Delta Y/\Delta Z$ much higher than the standard one 
(i.e. $\Delta Y/\Delta Z = 5$ instead of 2). Moreover, we checked the systematic differences in age estimates due to the inclusion or not of the microscopic element diffusion 
in stellar evolution calculations, which is an important bias source in grid-based techniques \citep{scepter1, eta}. Finally, we quantified  the effect of changing the mixing-length 
calibration in stellar models. 
 
The structure of the paper is the following. In Sect.~\ref{sec:methods} we 
discuss the estimation method and the stellar-model grids used in the process. 
The age estimates are presented in Sect.~\ref{sec:estimates},  and the analysis of the impact of changing the 
grid of models by taking into account a much higher helium abundance, by neglecting the element microscopic diffusion, and by increasing the 
mixing-length parameter is performed in Sect.~\ref{sec:bias-sources}. Some
concluding remarks can be found in Sect.~\ref{sec:conclusions}.
Finally, a Monte Carlo validation of the adopted method for stars in evolutionary stages similar to those of the microlensed sample is presented in Appendix~\ref{sec:bias}. 
 
\section{Methods}\label{sec:methods}

The age estimates obtained in this paper are based on our recently developed pipeline SCEPtER, 
which is extensively described in \citep{scepter1, eta} and briefly summarized  here. 
We let $\cal S$ be a star for which the following vector of observed quantities
is available: $q^{\cal S} \equiv \{T_{\rm eff, \cal S}, {\rm [Fe/H]}_{\cal S},
\log g_{\cal S}\}$. Then we let $\sigma = \{\sigma(T_{\rm
        eff, \cal S}), \sigma({\rm [Fe/H]}_{\cal S}), \sigma(\log g_{\cal S})\}$ 
be the nominal uncertainty in the observed
quantities. For each point $j$ on the estimation grid of stellar models, 
we define $q^{j} \equiv \{T_{{\rm eff}, j}, {\rm [Fe/H]}_{j}, \log g_{j}\}$. 
We let $ {\cal L}_j $ be the likelihood function defined as
\begin{equation}
        {\cal L}_j = \left( \prod_{i=1}^3 \frac{1}{\sqrt{2 \pi} \sigma_i} \right)
        \times \exp \left( -\frac{\chi^2}{2} \right)
        \label{eq:lik}
,\end{equation}
where
\begin{equation}
        \chi^2 = \sum_{i=1}^3 \left( \frac{q_i^{\cal S} - q_i^j}{\sigma_i} \right)^2.
\end{equation}

The likelihood function is evaluated for each grid point within $3 \sigma$ of
all the variables from $\cal S$; we let ${\cal L}_{\rm max}$ be the maximum value
obtained in this step. The estimated stellar mass, radius,
and age are obtained
by averaging the corresponding quantity of all the models with likelihood
greater than $0.95 \times {\cal L}_{\rm max}$.

The technique allows  a Monte Carlo confidence
interval to be constructed for the estimates. 
To this purpose a synthetic sample of $n$ stars is
generated, following a multivariate normal distribution with vector of mean
$q^{\cal S}$ and covariance matrix $\Sigma = {\rm diag}(\sigma)$. A value of
$n = 10\,000$ is usually adopted since it provides a fair balance between
computation time and the accuracy of the results.  The medians of the estimated 
mass and age for th $n$ objects are taken as the best estimate of the true values;
the 
16th and 84th quantiles of the $n$ values are adopted as a $1 \sigma$
confidence interval. 

\subsection{Standard stellar models grid}\label{sec:modelgrid}

The standard estimation grid of stellar models was obtained using the FRANEC
stellar evolution code \citep{scilla2008, tognelli2011}, as it was adopted to compute the Pisa Stellar
Evolution Data Base\footnote{\url{http://astro.df.unipi.it/stellar-models/}} 
for low-mass stars \citep{database2012, stellar}.

The grid consists of stellar models 
in evolutionary stages from the zero-age main sequence (ZAMS) to the helium flash at the tip of the red giant branch (RGB).
Models of age greater than 15 Gyr were excluded from the grid. Models were computed for stellar masses in the range [0.60; 1.60]
$M_{\sun}$ with a step of 0.02 $M_{\sun}$. The initial metallicity [Fe/H]
was assumed in the range [$-2.0$; 0.55] with a step of 0.10 dex from $-2.0$ to $-1.0$ and 0.05 dex for [Fe/H] > $-1.0$. The solar scaled heavy-element mixture by
\citet{AGSS09} was adopted. For [Fe/H] lower than $-0.5$ dex we accounted for an $\alpha$-elements enhancement [$\alpha$/Fe] = $+0.3$.
 The initial helium abundance was obtained using the
linear relation $Y = Y_p+\frac{\Delta Y}{\Delta Z} Z$.
The primordial $^4$He abundance value $Y_p = 0.2485$ from WMAP was adopted
\citep{cyburt04,steigman06,peimbert07a,peimbert07b}, assuming $\Delta
Y/\Delta Z = 2$ \citep{pagel98,jimenez03,gennaro10}. The models were computed
assuming our solar-scaled mixing-length parameter $\alpha_{\rm
        ml} = 1.74$. Atomic diffusion was included adopting the coefficients given by
\citet{thoul94} for gravitational settling and thermal diffusion. Convective core overshooting was
not taken into account.
Further details on the input of the code can be found in \citet{cefeidi,incertezze1,incertezze2}.

A Monte Carlo estimation of the expected errors and possible biases of our technique for stars similar to those in the microlensed bulge sample, adopting the available set of observational constraints, is reported in Appendix~\ref{sec:bias}. The analysis showed that the method is unbiased for the considered evolutionary phases. 

The adopted estimation technique and the grid of stellar models differ under many aspects from those used by \citet{Bensby2013}, therefore a confirmation of their finding would add considerable robustness to the age results. 
\citet{Bensby2011,Bensby2013} adopted a maximum likelihood functional maximization, where the errors on the estimates are derived from the likelihood profiles. They used the stellar models by Yonsei-Yale isochrones \citep{Demarque2004}, which rely on input physics different from our own reference. In particular, they adopt the solar mixture from \citet{Grevesse1996}, the low temperatures opacities by \citet{ferg94}, the conductive opacities by \citet{hubbard69}, the OPAL equation of state in the 1996 version \citep{rogers1996}, helium but no metal microscopic diffusion, and the cross sections listed in \citet{Bahcall1989}. Moreover, they take into account a mild convective core overshooting following the recipe described in \citet[][]{Demarque2004}.

\section{Age estimates}\label{sec:estimates}

\begin{figure}
        \centering
        \includegraphics[height=8.5cm,angle=-90]{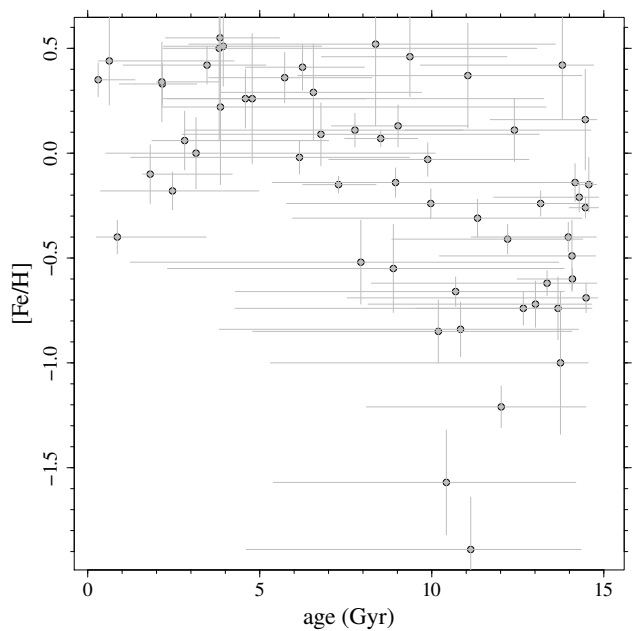}
        \caption{[Fe/H] versus SCEPtER age estimates for the 58 microlensed stars.} 
        \label{fig:agefeh}
\end{figure}

\begin{figure*}
\centering
\includegraphics[height=8.5cm,angle=-90]{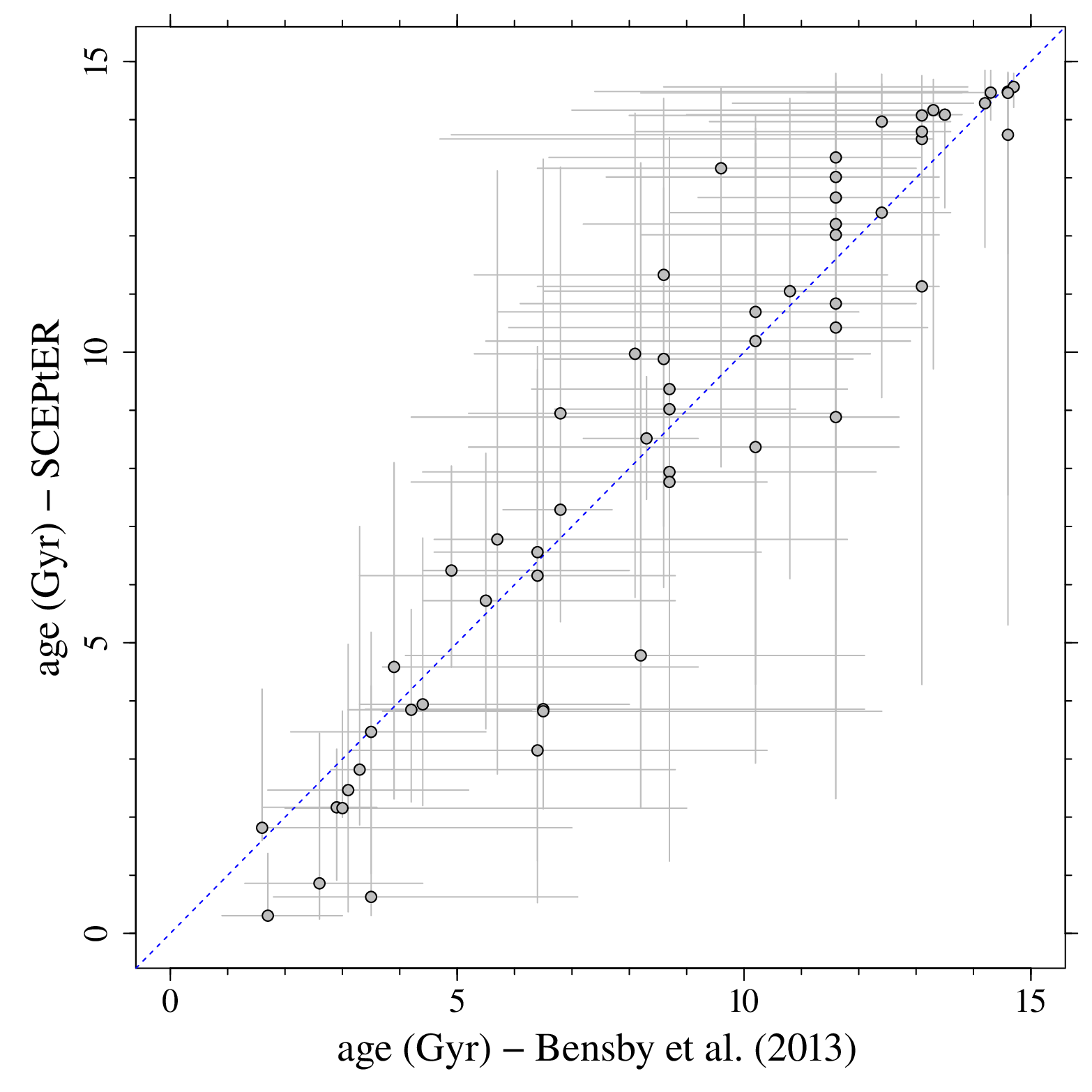}
\includegraphics[height=8.5cm,angle=-90]{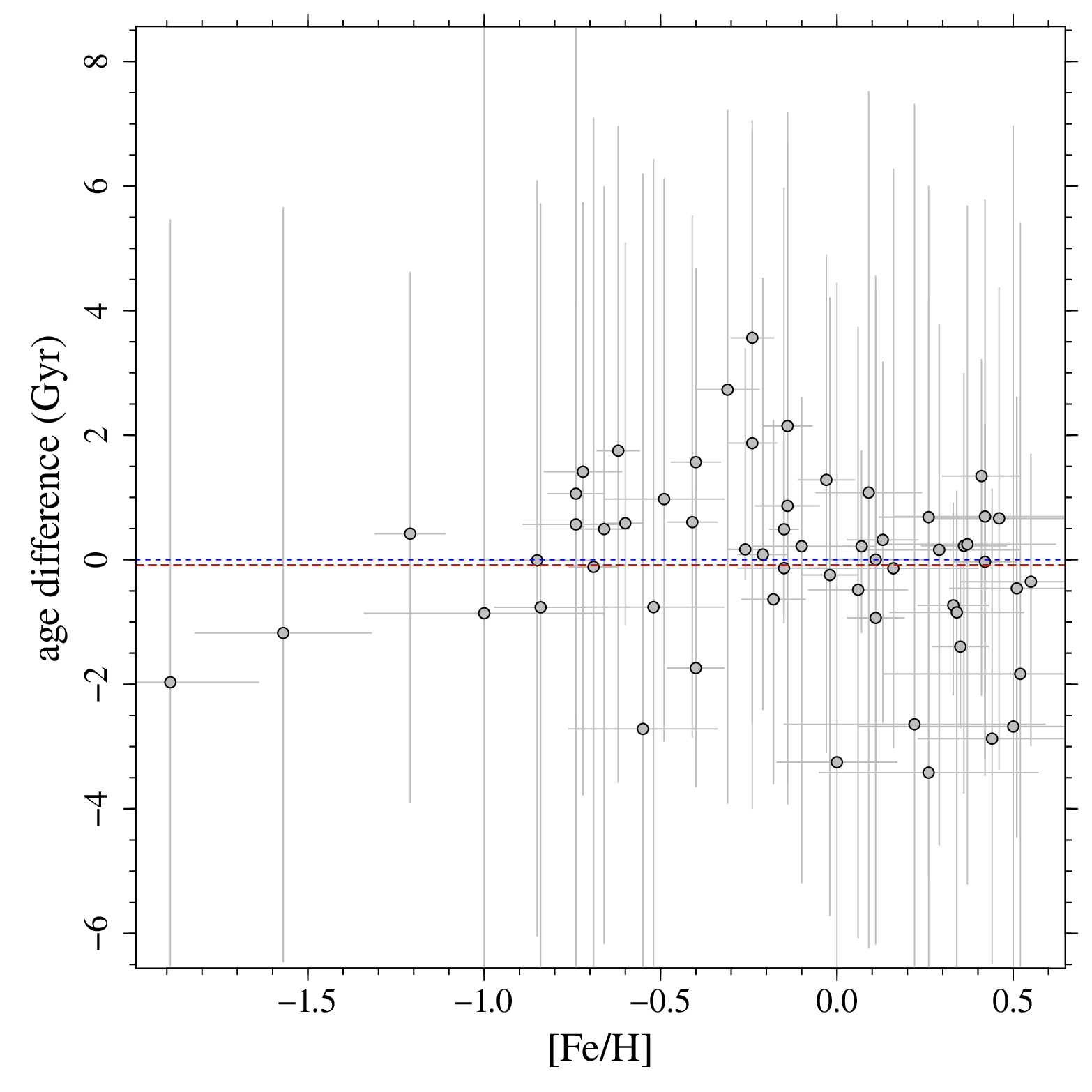}
\caption{{\it Left}: SCEPtER age estimates for the 58 microlensed stars versus the \citet{Bensby2013} estimates. {\it Right}: difference in age estimates between the two techniques versus [Fe/H]. A positive value means that the SCEPtER age is greater than the corresponding age estimated by \citet{Bensby2013}. The red dashed line marks the mean difference of $-0.08$ Gyr between the two sets of age estimates.} 
\label{fig:age-cfr}
\end{figure*}

The observational data -- with their errors -- of the 58 microlensed bulge stars studied in this paper have been obtained from Table~4 in \citet{Bensby2013}. For each object, we estimated the mass and the age by means of the SCEPtER technique, as detailed in Sect.~\ref{sec:methods}.
Figure~\ref{fig:agefeh} and Tab.~\ref{tab:estimates} show our results together with the observed [Fe/H]. 
Figure~\ref{fig:age-cfr} shows the comparison between our age estimates and those by \citet{Bensby2013} (left panel), and the age differences between the former and the latter as a function of the metallicity (right panel). 
The error bars on the age differences are obtained by adding in quadrature the errors provided by the two techniques. As one can see, the two independent results generally agree with each other, as it is apparent that all the differences are consistent with zero. 
A comparison of the two sets  by means of a paired $t$-test  showed a small mean difference of $-0.08$ Gyr (95\% confidence interval [$-0.46$; 0.30] Gyr), visible in Fig.~\ref{fig:age-cfr}. 

\begin{table*}[ht]
        \centering
        \caption{Ages and masses determined by the SCEPtER pipeline.}
        \label{tab:estimates}   
        \begin{tabular}{lcccccccc}
        \hline\hline
        Object             &  age  & $-1 \sigma$ & $1 \sigma$ &     mass     & $-1 \sigma$  &  $1 \sigma$  & [Fe/H] & $\Delta$ [Fe/H]\\
        & (Gyr) &    (Gyr)    &   (Gyr)    & ($M_{\sun}$) & ($M_{\sun}$) & ($M_{\sun}$) & & \\ \hline
 OGLE-2012-BLG-1156 & 11.1 & 6.5 & 3.2 & 0.76 & 0.05 & 0.09 & -1.89 & 0.25 \\ OGLE-2011-BLG-0969 & 10.4 & 5.0 & 3.8 & 0.79 & 0.05 & 0.13 & -1.57 & 0.25 \\ MOA-2010-BLG-285 & 12.0 & 3.9 & 2.5 & 0.79 & 0.04 & 0.07 & -1.21 & 0.10 \\ MOA-2010-BLG-078 & 13.7 & 8.4 & 0.8 & 0.82 & 0.04 & 0.25 & -1.00 & 0.34 \\ MOA-2011-BLG-104 & 10.2 & 5.4 & 3.9 & 0.83 & 0.05 & 0.13 & -0.85 & 0.15 \\ OGLE-2012-BLG-0270 & 10.8 & 7.0 & 3.4 & 0.81 & 0.04 & 0.07 & -0.84 & 0.13 \\ MOA-2012-BLG-187 & 12.7 & 3.1 & 1.9 & 0.81 & 0.03 & 0.05 & -0.74 & 0.08 \\ MOA-2009-BLG-493 & 13.7 & 9.4 & 1.0 & 0.72 & 0.04 & 0.08 & -0.74 & 0.15 \\ OGLE-2009-BLG-076 & 13.0 & 4.9 & 1.6 & 0.80 & 0.04 & 0.04 & -0.72 & 0.11 \\ MOA-2009-BLG-133 & 14.5 & 6.9 & 0.3 & 0.74 & 0.02 & 0.06 & -0.69 & 0.07 \\ OGLE-2012-BLG-0563 & 10.7 & 6.4 & 3.2 & 0.81 & 0.03 & 0.04 & -0.66 & 0.07 \\ OGLE-2012-BLG-1279 & 13.4 & 5.1 & 1.4 & 0.78 & 0.02 & 0.03 & -0.62 & 0.06 \\ MOA-2010-BLG-167 & 14.1 & 1.6 & 0.1 & 0.82 & 0.00 & 0.03 & -0.60 & 0.05 \\ MOA-2012-BLG-532 & 8.9 & 6.6 & 5.0 & 0.90 & 0.10 & 0.45 & -0.55 & 0.21 \\ MOA-2009-BLG-475 & 7.9 & 6.7 & 5.8 & 0.84 & 0.06 & 0.08 & -0.52 & 0.20 \\ MACH-1999-BLG-022 & 14.1 & 3.9 & 0.7 & 0.83 & 0.04 & 0.08 & -0.49 & 0.17 \\ OGLE-2012-BLG-1217 & 12.2 & 3.4 & 2.2 & 0.83 & 0.03 & 0.04 & -0.41 & 0.07 \\ MOA-2010-BLG-049 & 14.0 & 2.8 & 0.8 & 0.84 & 0.02 & 0.06 & -0.40 & 0.07 \\ MOA-2010-BLG-446 & 0.9 & 0.6 & 2.6 & 1.01 & 0.05 & 0.05 & -0.40 & 0.08 \\ OGLE-2008-BLG-209 & 11.3 & 5.4 & 3.0 & 0.90 & 0.06 & 0.18 & -0.31 & 0.09 \\ MOA-2011-BLG-090 & 14.5 & 0.5 & 0.4 & 0.86 & 0.02 & 0.00 & -0.26 & 0.05 \\ OGLE-2012-BLG-1526 & 13.2 & 5.1 & 1.4 & 0.88 & 0.02 & 0.13 & -0.24 & 0.06 \\ MOA-2012-BLG-391 & 10.0 & 4.2 & 4.1 & 0.95 & 0.09 & 0.15 & -0.24 & 0.07 \\ MOA-2009-BLG-489 & 14.3 & 2.5 & 0.6 & 0.86 & 0.02 & 0.05 & -0.21 & 0.07 \\ MOA-2011-BLG-174 & 2.5 & 2.1 & 2.5 & 1.05 & 0.05 & 0.07 & -0.18 & 0.09 \\ OGLE-2012-BLG-1534 & 7.3 & 1.0 & 1.1 & 1.04 & 0.05 & 0.05 & -0.15 & 0.04 \\ MOA-2012-BLG-202 & 14.6 & 0.4 & 0.2 & 0.88 & 0.02 & 0.02 & -0.15 & 0.13 \\ OGLE-2012-BLG-0617 & 14.2 & 4.5 & 0.5 & 0.89 & 0.03 & 0.10 & -0.14 & 0.09 \\ MOA-2012-BLG-410 & 8.9 & 3.6 & 4.2 & 0.99 & 0.10 & 0.15 & -0.14 & 0.07 \\ OGLE-2012-BLG-0816 & 1.8 & 0.2 & 2.4 & 1.57 & 0.34 & 0.03 & -0.10 & 0.14 \\ OGLE-2012-BLG-0211 & 9.9 & 2.9 & 2.9 & 0.99 & 0.08 & 0.10 & -0.03 & 0.08 \\ MOA-2011-BLG-234 & 6.2 & 4.9 & 3.2 & 0.96 & 0.04 & 0.05 & -0.02 & 0.08 \\ OGLE-2011-BLG-1105 & 3.1 & 2.6 & 6.9 & 0.93 & 0.06 & 0.07 & 0.00 & 0.17 \\ MOA-2010-BLG-523 & 2.8 & 1.0 & 4.2 & 1.42 & 0.33 & 0.17 & 0.06 & 0.14 \\ OGLE-2012-BLG-1274 & 8.5 & 1.0 & 1.1 & 1.03 & 0.04 & 0.04 & 0.07 & 0.04 \\ OGLE-2012-BLG-0521 & 6.8 & 4.0 & 6.3 & 1.12 & 0.19 & 0.33 & 0.09 & 0.15 \\ MOA-2011-BLG-034 & 12.4 & 3.2 & 2.2 & 0.94 & 0.05 & 0.09 & 0.11 & 0.15 \\ MOA-2009-BLG-174 & 7.8 & 5.0 & 3.1 & 0.94 & 0.04 & 0.04 & 0.11 & 0.08 \\ MOA-2009-BLG-456 & 9.0 & 1.9 & 2.0 & 1.00 & 0.06 & 0.06 & 0.13 & 0.10 \\ MOA-2012-BLG-291 & 14.5 & 2.8 & 0.3 & 0.92 & 0.06 & 0.06 & 0.16 & 0.24 \\ OGLE-2011-BLG-1410 & 3.9 & 1.7 & 9.5 & 1.35 & 0.41 & 0.23 & 0.22 & 0.37 \\ MOA-2011-BLG-191 & 4.6 & 2.3 & 3.5 & 1.27 & 0.18 & 0.26 & 0.26 & 0.14 \\ MOA-2011-BLG-445 & 4.8 & 2.6 & 8.5 & 1.26 & 0.32 & 0.31 & 0.26 & 0.31 \\ OGLE-2007-BLG-514 & 6.6 & 2.7 & 3.1 & 1.09 & 0.11 & 0.16 & 0.29 & 0.23 \\ OGLE-2011-BLG-0950 & 2.2 & 1.3 & 1.0 & 1.29 & 0.09 & 0.12 & 0.33 & 0.10 \\ MOA-2009-BLG-259 & 2.2 & 0.2 & 1.7 & 1.59 & 0.24 & 0.01 & 0.34 & 0.19 \\ MOA-2008-BLG-311 & 0.3 & 0.1 & 1.1 & 1.14 & 0.04 & 0.04 & 0.35 & 0.08 \\ OGLE-2011-BLG-1072 & 5.7 & 2.2 & 2.5 & 1.20 & 0.12 & 0.17 & 0.36 & 0.12 \\ MOA-2011-BLG-058 & 11.0 & 4.9 & 3.3 & 1.00 & 0.08 & 0.18 & 0.37 & 0.25 \\ MOA-2008-BLG-310 & 6.2 & 1.7 & 1.8 & 1.09 & 0.06 & 0.07 & 0.41 & 0.11 \\ MOA-2012-BLG-022 & 3.5 & 2.4 & 1.7 & 1.15 & 0.06 & 0.07 & 0.42 & 0.09 \\ OGLE-2007-BLG-349 & 13.8 & 4.1 & 0.9 & 0.93 & 0.05 & 0.06 & 0.42 & 0.26 \\ MOA-2006-BLG-099 & 0.6 & 0.3 & 3.6 & 1.08 & 0.07 & 0.06 & 0.44 & 0.21 \\ OGLE-2006-BLG-265 & 9.4 & 2.6 & 2.8 & 1.00 & 0.06 & 0.07 & 0.46 & 0.19 \\ OGLE-2012-BLG-0026 & 3.8 & 1.6 & 9.2 & 1.37 & 0.41 & 0.21 & 0.50 & 0.44 \\ MOA-2010-BLG-311 & 3.9 & 1.7 & 2.9 & 1.33 & 0.19 & 0.24 & 0.51 & 0.19 \\ MOA-2011-BLG-278 & 8.4 & 5.4 & 5.2 & 1.04 & 0.12 & 0.33 & 0.52 & 0.39 \\ MOA-2010-BLG-037 & 3.8 & 1.6 & 1.7 & 1.31 & 0.14 & 0.23 & 0.55 & 0.20 \\
 \hline
        \end{tabular}
\tablefoot{Column 1: star identifier;  Col. 2: stellar age;  Cols. 3 and 4: $1 \sigma$ lower and upper uncertainties for age estimates; Col. 5: stellar mass; Cols. 6 and 7: $1 \sigma$ lower and upper uncertainties for mass estimates;  Cols. 8 and 9: [Fe/H] and its error from \citet{Bensby2013}.}
\end{table*}

\begin{figure}
\centering
\includegraphics[height=8.5cm,angle=-90]{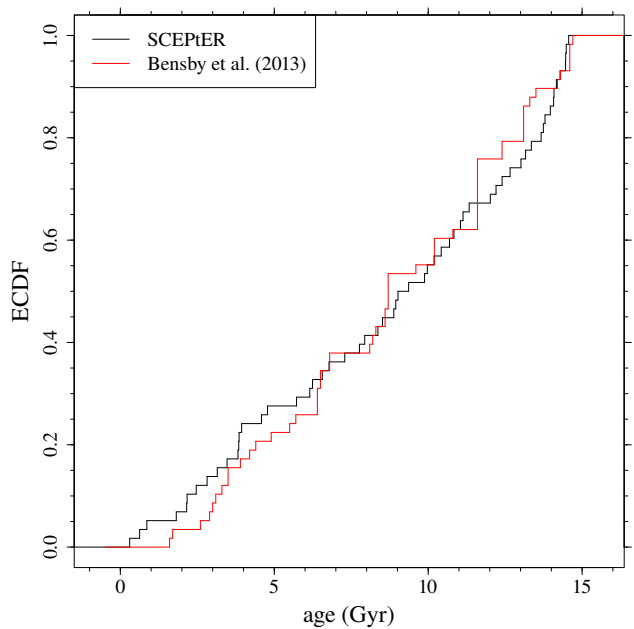}
\caption{Empirical cumulative distribution functions for the 58 estimated ages obtained by SCEPtER (black line) and by \citet{Bensby2013} (red line). 
        } 
\label{fig:age-ecdf-feh}
\end{figure}

This agreement is further confirmed by comparing the empirical cumulative distributions of stellar ages estimated by the two techniques, shown in Fig.~\ref{fig:age-ecdf-feh}. 
A two-sample Kolmogorov-Smirnov test did not reveal significant differences between the two empirical distributions ($D = 0.105$, $p$-value = 0.92) implying that the age distribution can be considered consistent between the two estimation methods. 

It is apparent from Fig.~\ref{fig:age-ecdf-feh} that both the techniques independently estimate a sizeable fraction of young stars. We found 16 stars younger than 5 Gyr and 28 younger than 9 Gyr that is,
  28\% (95\% confidence interval [17\%; 41\%]) and  48\%  (95\% confidence interval [35\%; 62\%]) of the sample, respectively, whereas \citet{Bensby2013}  found 13 (22\%) and 31 (58\%).
   
\section{Effect of changing the grid of stellar models}\label{sec:bias-sources}

The determination of stellar ages necessarily requires theoretical stellar models, which in turn depend on  prescriptions and parameters that have not yet been tightly constrained \citep{incertezze1, incertezze2}.
As a consequence, the resulting age estimates are affected by non-negligible uncertainties and, possibly, biases. In  recent years, several studies have been devoted to the analysis of the performance of 
grid based techniques \citep[e.g.][]{Gai2011, Basu2012, Mathur2012}. In this perspective, in \citet{scepter1, eta} we carried out a detailed statistical analysis of the main biases affecting any grid-based 
recovery procedure. In this section we will discuss the impact on the age estimate of the 58 microlensed bulge stars of  increasing the initial helium abundance, switching off microscopic diffusion, and 
changing the mixing-length calibration adopted to compute the stellar models. The main aim of the test is to check the robustness of the assessment of a sizeable fraction of young bulge stars.

\subsection{Effect of increasing the initial helium abundance}\label{sec:bias-he}

\citet{Nataf2012} suggested that the discrepancy in age 
between \citet{Bensby2013} results and those inferred by photometric data
would vanish whenever an enhancement in the helium content of the bulge stars was addressed. \citet{Nataf2012}, studying the effect on theoretical isochrones of helium abundance variations, discussed the systematic effects on age estimates that could arise when observational data were analysed with standard instead of helium-enhanced isochrones. 
In particular, they found that the predicted trend of the difference between the true absolute magnitude and the spectroscopically inferred one should be positively correlated with the spectroscopically  inferred stellar mass, a trend that they found also present in the \citet{Bensby2011} data. Thus they claimed that a value of helium-to-metal enrichment ratio of $\Delta Y/\Delta Z \approx 5$ would reconcile the two sets of age estimates.
In contrast, \citet{Bensby2013} showed that the discussed trend is no longer present in the extended stellar sample; they rejected the hypothesis by \citet{Nataf2012} and concluded that a sizeable sample of young and intermediate age stars would still be present 
even assuming high helium abundances.  
However, neither \citet{Nataf2012} nor \citet{Bensby2013} performed a statistically detailed comparison of the age estimates obtained with different values of the original helium abundance.
Moreover, the impact of assuming a different initial helium value is strongly dependent on the observational quantities used to constrain the ages. As an example, in \citet{eta} we have recently shown that in the presence of asteroseismic constraints the impact on age estimates of a change by $\pm 1$ in $\Delta Y/\Delta Z$ is indeed negligible, while the contrary occurs if mass and radius are available  \citep{binary}.       

\begin{figure*}
\centering
\includegraphics[height=8.5cm,angle=-90]{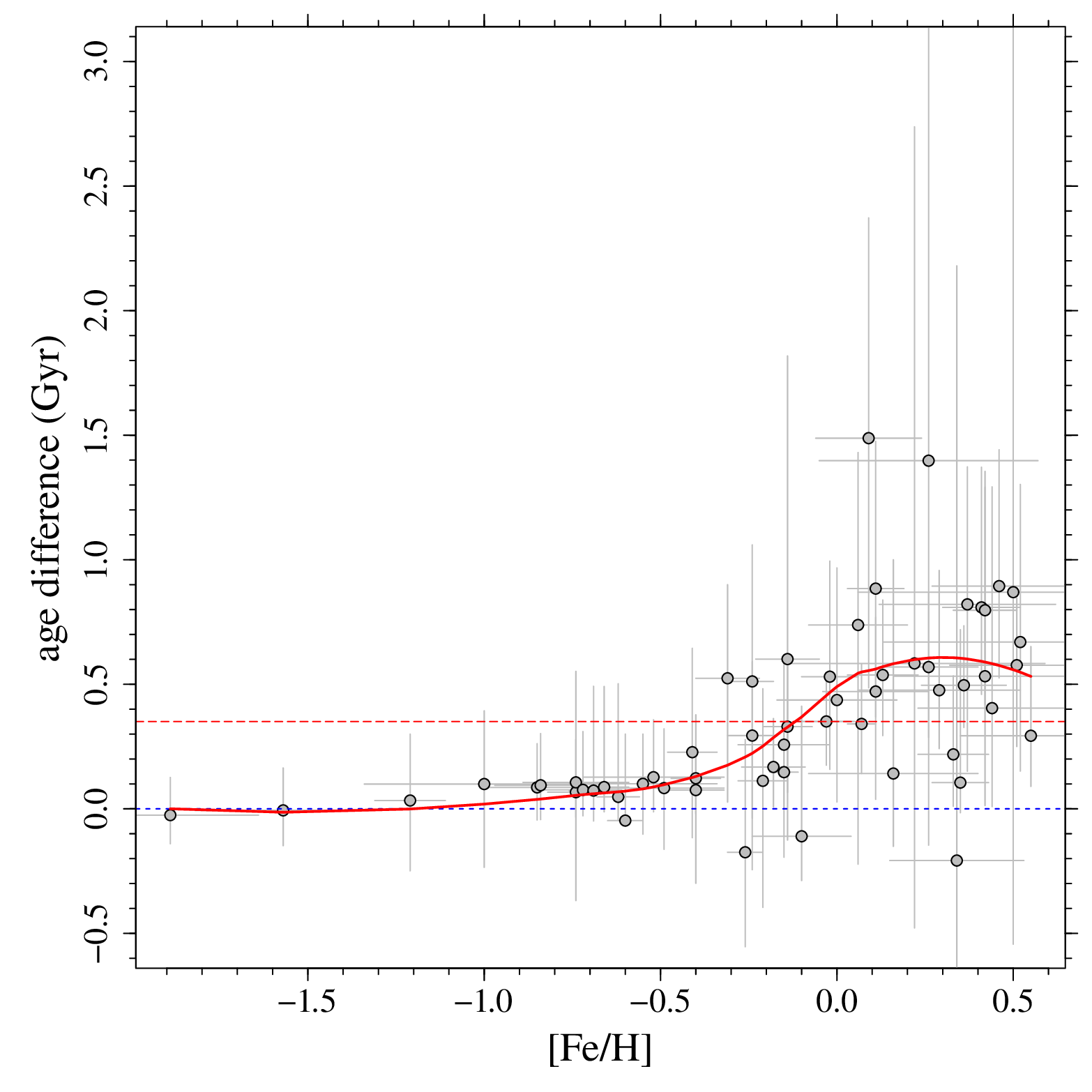}
\includegraphics[height=8.5cm,angle=-90]{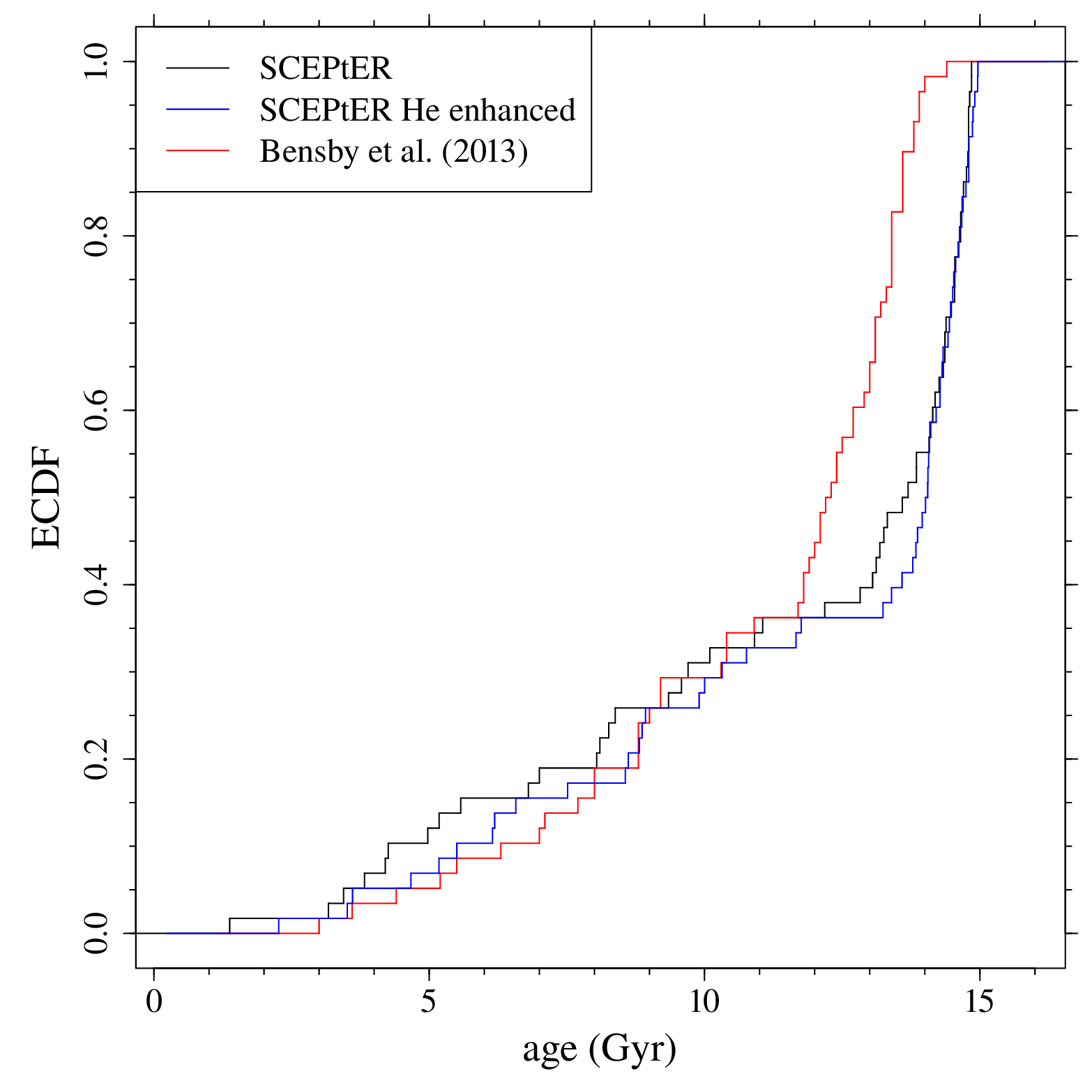}
\caption{{\it Left}: differences between the age estimates obtained by adopting the grid with $\Delta Y/\Delta Z = 5$ (helium-enhanced) and that with $\Delta Y/\Delta Z = 2$ (standard). The red dashed line marks the mean of the differences, while the solid red line is a LOESS smoother of the data (see text). {\it Right}: empirical cumulative distribution functions for the upper limits of the estimated ages obtained by SCEPtER assuming the standard initial helium value (black line), the helium-enhanced scenario (blue line), and for the ages estimated by \citet{Bensby2013} (red line).} 
\label{fig:cfrHe5}
\end{figure*}

Because of its potential impact in reducing the number of estimated young bulge stars, the  effect of a large increase in the initial helium content is worthy of a detailed quantitative analysis. 
Therefore, we computed a non-standard grid of models with the same parameters or input physics and the same values of the metallicity $Z$ as in the standard grid, but by significantly increasing the helium-to-metal enrichment ratio, i.e. $\Delta Y/\Delta Z = 5$ instead of 2. Then we repeated the age estimate by adopting this helium-enhanced grid in the SCEPtER pipeline. To account for possible correlations among the estimates given by the standard and helium-enhanced grids, some care was needed in the comparison. 
In fact, the naive differences of the two estimates (and their errors) could be obtained
by subtracting the estimates provided by the standard and helium-enhanced grids and by adding in quadrature the errors in the single estimates. However, since such an approach largely overestimates the final errors, we preferred to follow a more accurate procedure. 
For each star, we obtained $N = 10\,000$ artificially perturbed stars adding  a Gaussian noise to the observables as detailed in Sect.~\ref{sec:methods}. The artificial star ages are then estimated by means of the SCEPtER pipeline with both grids. For each of the  $N$ artificial stars, we evaluated the difference between the age returned by the helium-enhanced and the standard grids. The median of these differences was adopted as best estimate and the 16th and 84th quantiles were adopted as $1 \sigma$ error.  

The results of this analysis are given in Fig.~\ref{fig:cfrHe5}. The left panel of the figure shows the differences in the estimates computed with the helium-enhanced and the standard grid of models. It appears that the change in the initial helium content has a little influence in age estimates. Adopting the helium-enhanced grid, the mean increase in age estimates  is 0.35 Gyr (95\% confidence interval [0.26; 0.44] Gyr). As expected, the influence of the initial helium abundance change is greater at high metallicity. The effect is shown in the figure by superimposing  a LOESS smoother\footnote{A LOESS (LOcal regrESSion) 
        smoother is 
        a non-parametric  locally weighted polynomial regression technique that is
        often used to highlight the underlying trend of scattered data \citep[see
        e.g.][]{Feigelson2012,venables2002modern}.} on the data: the higher the metallicity, the larger the discrepancy, until a plateau of about 0.6 Gyr for [Fe/H] > 0. However, such an increase in the age estimates 
        is too small to rule out the evidence of a young component in the microlensed population. 

As one can see in Fig.~\ref{fig:age-cfr}, the error bars on the age estimates are relevant. An important point to address is thus whether the previous finding might result simply from a fluctuation. 
Since we are mainly interested in checking the robustness of the identification of a sizeable fraction of young stars, we further analysed our results by 
considering as age estimates those provided by the upper values of the error bars 
 rather than the mean values. With such a choice, the probability of obtaining 
by chance an age older than the adopted one is  16\%. 
The right panel in Fig.~\ref{fig:cfrHe5} shows the empirical cumulative density function for the upper values obtained by SCEPtER assuming standard and helium-enhanced grids of models, and by \citet{Bensby2013}. The three cumulative distributions closely agree up to 12 Gyr.
All  three sets of estimates contain 15 stars younger than 9 Gyr (26\% of the sample, 95\% confidence interval [16\%; 39\%]).
For stars younger than 5 Gyr, the estimates assuming standard helium abundance show the presence of seven  such objects (12\% of the sample, 95\% confidence interval [5\%; 24\%]), while there are four (7\% of the sample, 95\% confidence interval [2\%; 18\%]) in the helium-enhanced computations.  

Thus, even in the most conservative scenario in which we contemporaneously adopt a high initial helium abundance and the upper values of the age estimates, there is  strong evidence of a young population among the microlensed bulge stars. 

\subsection{Effect of neglecting microscopic diffusion}\label{sec:bias-diffusion}

An  important source of systematic bias in the age estimate of dwarf stars is the treatment of element microscopic diffusion adopted in stellar model computations. Depending on its efficiency, this process affects the evolution of both the surface chemical abundance  -- and thus [Fe/H] -- 
and the internal structure of low-mass stars. As a direct consequence, the stellar parameters determined by comparing theoretical models and observations depend on the diffusion efficiency adopted. In \citet{scepter1, eta} we showed that grid-based estimates of mass, radius, and age are 
significantly different between grids of models that take into account diffusion and those that neglect it. From the observational point of view, microscopic diffusion has been proved to be efficient in the Sun \citep[see e.g.][]{Bahcall2001}, while its actual efficiency in Galactic globular cluster stars is  debated \citep[see e.g.][]{Gruyters2014, Nordlander2012, Gratton2011, Korn2007}.

The \citet{Bensby2013} results are based on stellar models that consider only the diffusion of helium and not that of metals. On the contrary, our standard grid follows the diffusion of helium and metals, as mentioned in Sect.~\ref{sec:modelgrid}. Since our main aim is to verify whether the identification of young stars 
is robust against theoretical biases, we think that it is worth  exploring the effects of completely neglecting microscopic diffusion in stellar calculations. Such a choice would in fact maximize age estimates,  the other parameters being fixed, and consequently minimize the sample of young stars. We computed 
a non-standard grid of models with exactly the same characteristics as the standard model but without the inclusion of a diffusion mechanism. Then, we repeated the age estimate by adopting this non-standard grid in the SCEPtER pipeline. The differences between the age estimates with and without diffusion have 
been computed in the same way as discussed in Sect.~\ref{sec:bias-he} for the helium-rich case. 

\begin{figure*}
        \centering
        \includegraphics[height=8.5cm,angle=-90]{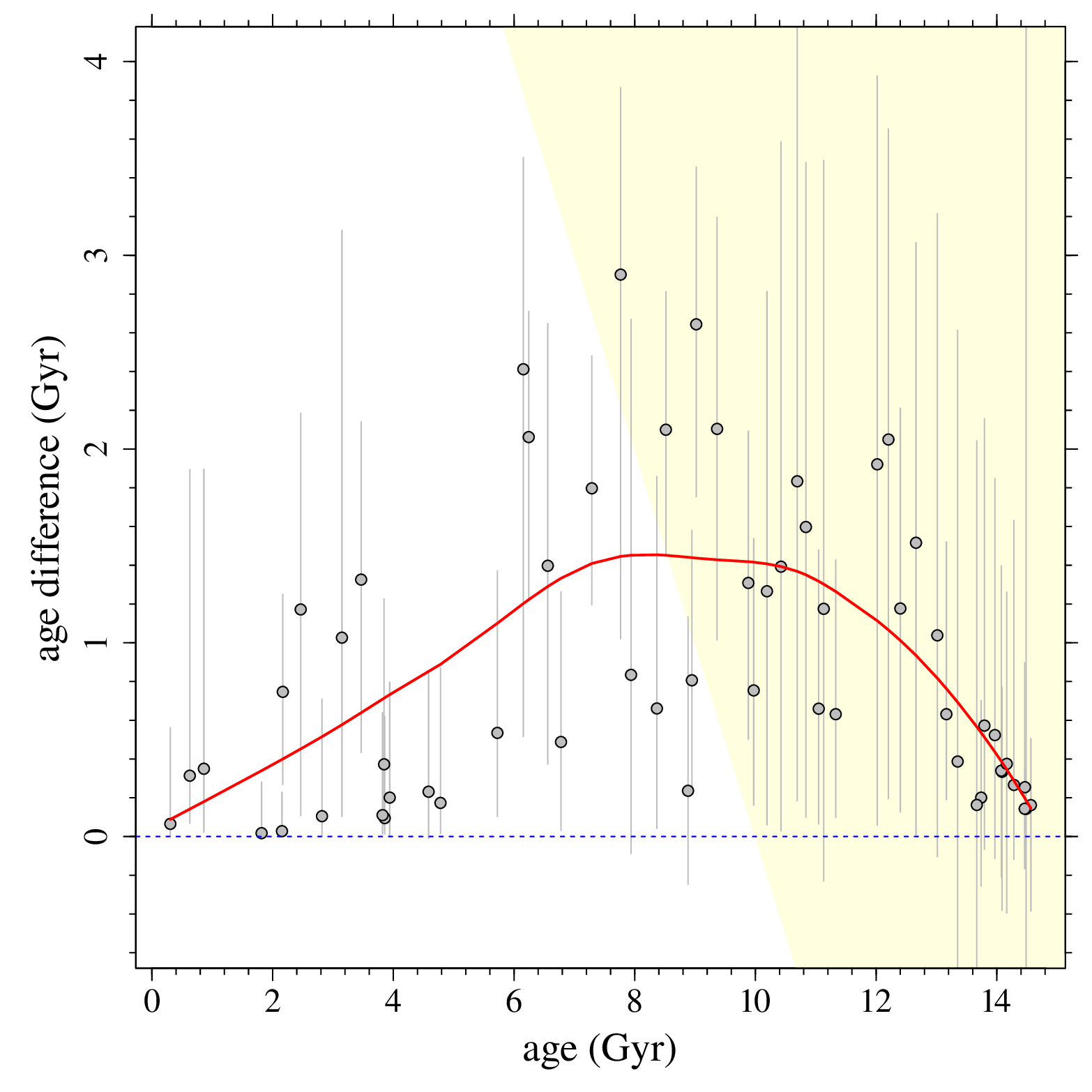}
        \includegraphics[height=8.5cm,angle=-90]{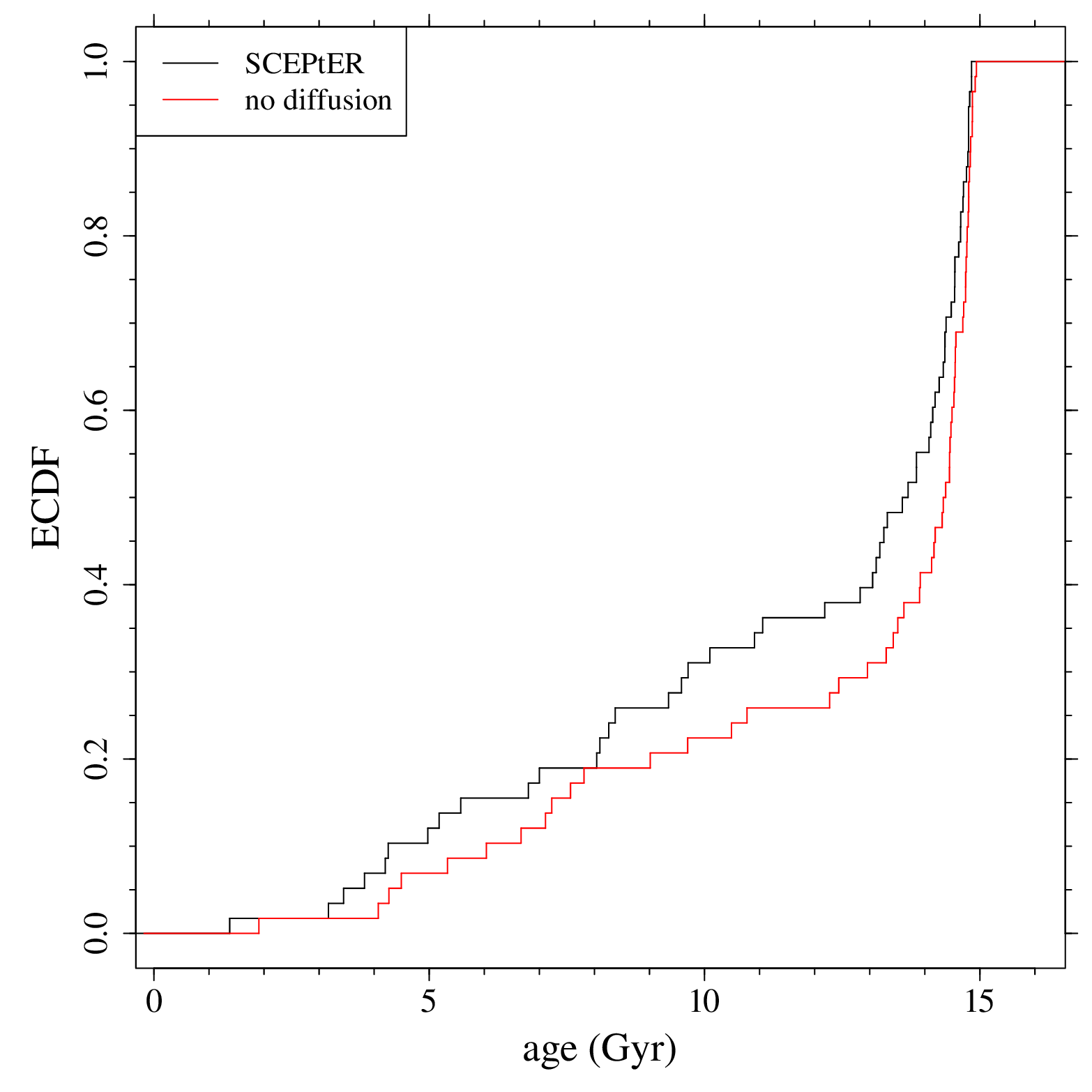} 
        \caption{{\it Left}: Differences in age estimates assuming the grid that neglects microscopic diffusion and the standard one. The solid red line is a LOESS smoother of the data (see text). The yellow region shows the stars older than 10 Gyr. {\it Right}: empirical cumulative distribution functions for the upper limits of the estimated ages obtained by SCEPtER adopting the grid with element diffusion (black line), and the grid that neglects it (red line).} 
        \label{fig:cfrdiff}
\end{figure*}

The results are shown in the left panel of Fig.~\ref{fig:cfrdiff}. As expected \citep[see e.g.][]{eta}, the neglect of diffusion in stellar models leads to older ages. However, such an effect gets smaller and smaller at decreasing ages owing to the long time scale of diffusion processes. 
The maximum increase of the estimated age is  about 2.8 Gyr at 8 Gyr and lower than 1.5 Gyr for stars younger than 6 Gyr. The mean age increase is indeed much smaller being of 0.40 Gyr (95\% confidence interval [0.17; 0.63] Gyr) for stars
younger than 5 Gyr. The coloured region in the figure highlights stars with estimated ages higher than 10 Gyr. 
The decreasing trend in the age differences for ages higher than about 10 Gyr is an edge-effect, a common feature of any grid-based technique, as extensively discussed in \citet{scepter1, eta}. Such an effect is the consequence of the lack of models older than 15 Gyr in the grid, which thus limits the possibility of age overestimation for these old stars.

The effect of the diffusion on age estimates can also be seen  in the right panel of Fig.~\ref{fig:cfrdiff}, which shows the empirical cumulative density functions for the upper values for the age estimates obtained considering or neglecting microscopic diffusion. 

It is apparent that even neglecting the microscopic diffusion into stellar model computations, there is no evidence against the conclusion of the presence of a relevant young stellar population. Even 
in the most unfavourable case in which the non-standard grid without diffusion and the upper values of the age estimates are used there are 4 (7\% of the sample, 95\% confidence interval [2\%; 18\%]) younger than 5 Gyr and  
11 (19\% of the sample, 95\% confidence interval [10\%; 32\%]) younger than 9 Gyr.   

\subsection{Effect of increasing the superadiabatic convection efficiency}\label{sec:bias-mixing}

Another source of systematic bias in age determination of dwarf stars is the efficiency of convective transport in superadiabatic regimes adopted in stellar computations. As is well known, modern evolutionary codes  still have to rely on oversimplified treatments of superadiabatic convection 
which depend on free parameters to be calibrated with observations. The most commonly adopted treatment is the mixing-length theory \citep{bohmvitense58}, where the efficiency of convective transport depends on the free parameter  $\alpha_{\rm ml}$, usually calibrated with the Sun. 
A solar calibrated mixing-length has been adopted both in our standard models and in those used by \citet{Bensby2013}. 
However, there is no a priori compelling reason to guarantee that a solar calibration also holds in evolutionary phases and/or masses different from that of the Sun. Moreover, there are indications that a slightly higher value ($\alpha_{\rm ml} = 1.90 $) than the solar calibrated is in better agreement with 
the colour-magnitude diagrams of old and intermediate age Galactic globular clusters \citep[see e.g.][]{database2012}. For this reason we think that it is worth discussing the effect on the age estimates of a change in the mixing-length value. As in the two previous tests, we focused only on a variation that would increase stellar ages, thus reducing the number of young bulge stars. 

\begin{figure*}
        \centering
        \includegraphics[height=8.5cm,angle=-90]{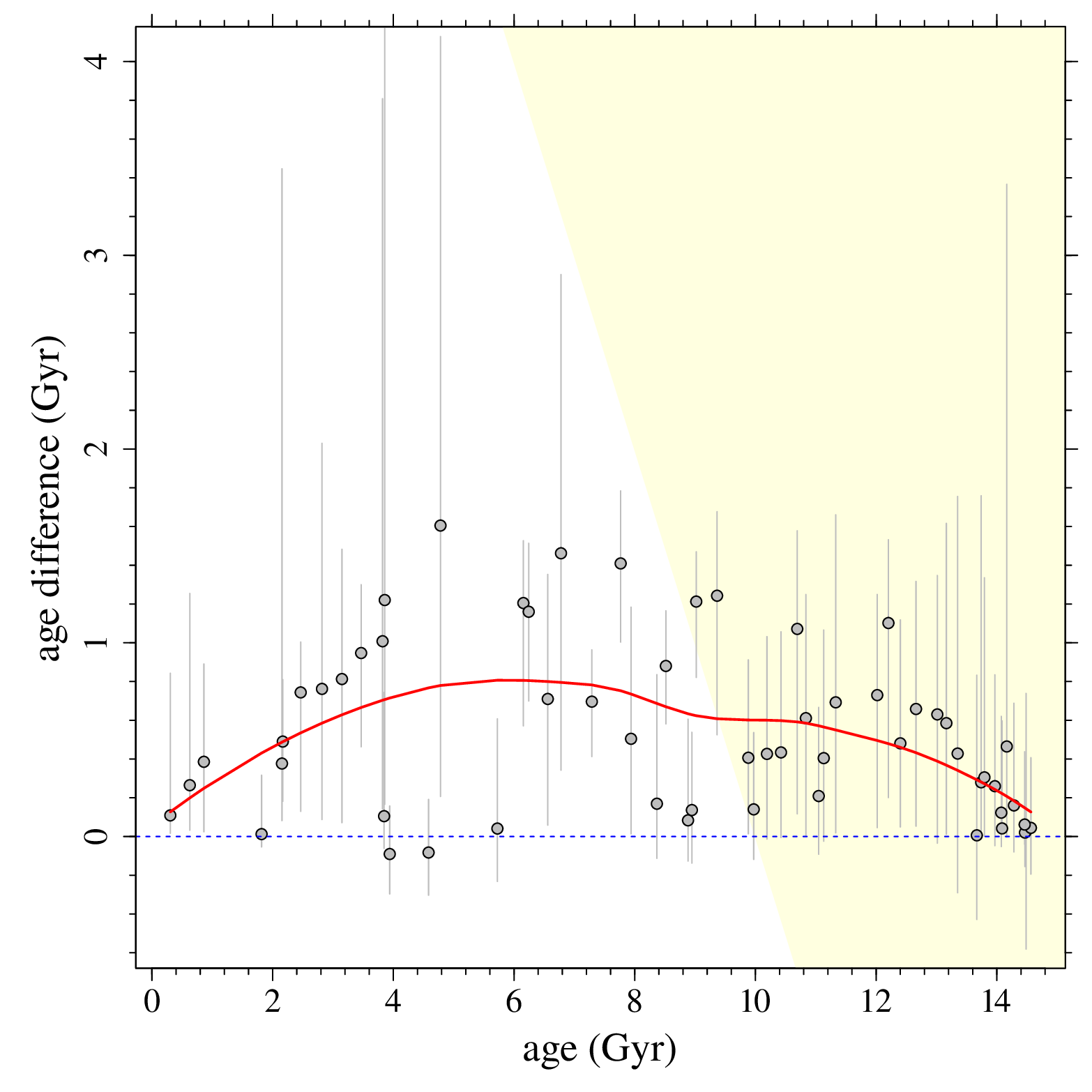}
        \includegraphics[height=8.5cm,angle=-90]{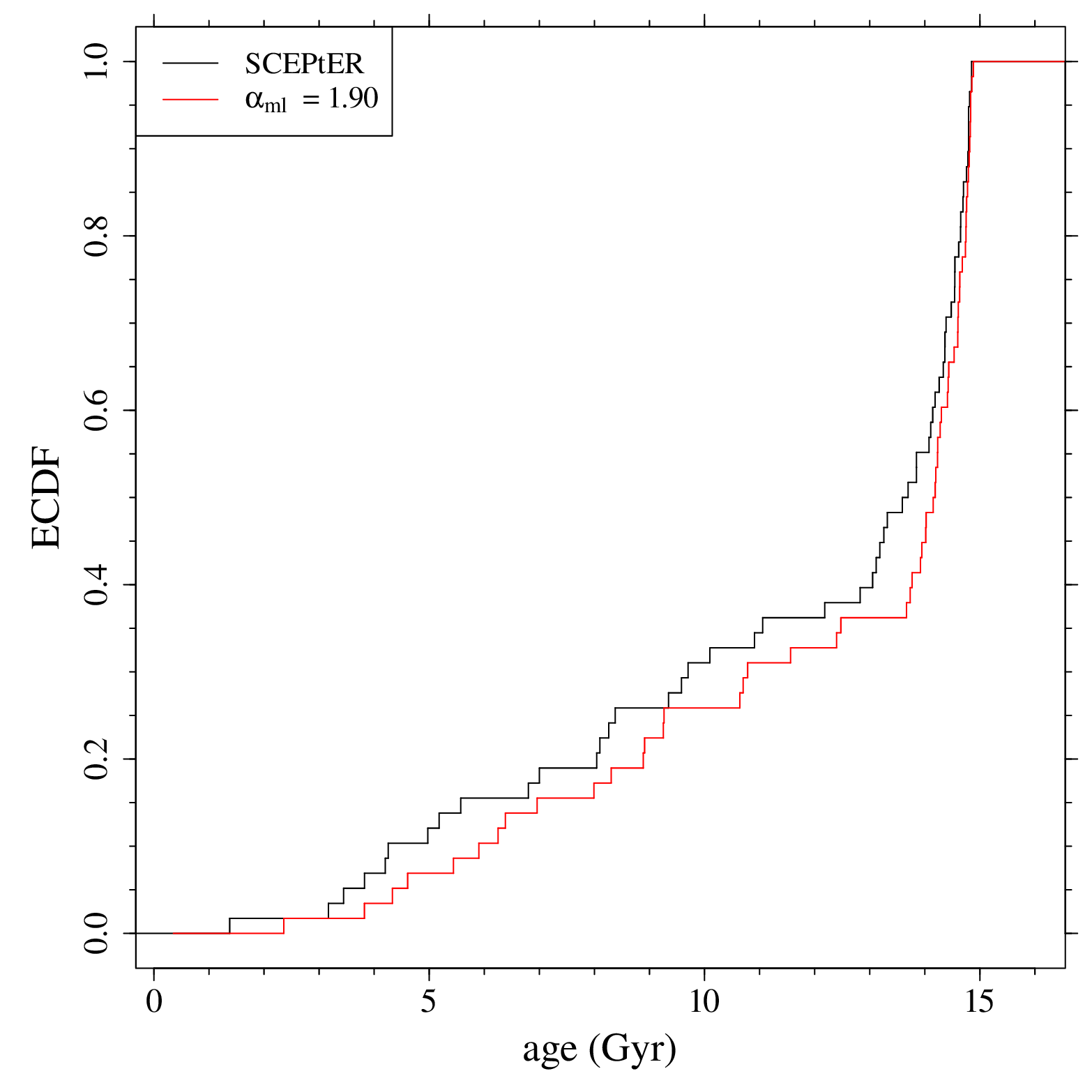}   
        \caption{{\it Left}: same as in the left panel of Fig.~\ref{fig:cfrdiff}, but for age differences due to assuming the grid with $\alpha_{\rm ml} = 1.90$ and the reference case with $\alpha_{\rm ml} = 1.74$. {\it Right}: empirical cumulative distribution functions for the upper limits of the estimated ages obtained by SCEPtER adopting the standard grid with $\alpha_{\rm ml} = 1.74$ (black line), and that with $\alpha_{\rm ml} = 1.90$ (red line). } 
        \label{fig:cfrML}
\end{figure*}

We computed a non-standard grid of models with the same characteristics of the standard one but computed with $\alpha_{\rm ml} = 1.90 $. Then, we repeated the age estimate by adopting this non-standard grid in the SCEPtER pipeline. The differences in ages between the $\alpha_{\rm ml} = 1.90 $ and the $\alpha_{\rm ml} = 1.74 $ (solar calibrated) cases have been computed as in the previous tests (see Sects.~\ref{sec:bias-he} and ~\ref{sec:bias-diffusion}). The results are presented in Fig.~\ref{fig:cfrML}. The age increase due to the higher mixing-length value is lower than the increase due to neglecting microscopic diffusion, and its maximum is about 1.5 Gyr. For stars younger than 5 Gyr, the mean increase in age estimate is 0.54 Gyr (95\% confidence interval [0.28; 0.81] Gyr).

The right panel of Fig.~\ref{fig:cfrML} shows the empirical cumulative density functions for the upper values of the age estimates obtained with grids of stellar models with the two considered values of $\alpha_{\rm ml}$. The differences between the two distributions are small. 
Even in the most unfavourable case, i.e.  simultaneously adopting the grid with $\alpha_{\rm ml} = 1.90 $ and the upper values of the age, estimates provide 4 (7\% of the sample, 95\% confidence interval [2\%; 18\%]) younger than 5 Gyr and  
13 (22\% of the sample, 95\% confidence interval [13\%; 36\%]) younger than 9 Gyr. 

In conclusion, even accounting for a higher mixing-length value, the presence of a relevant young stellar population is confirmed.      
  
\section{Conclusions}\label{sec:conclusions}

We independently verified the results by \citet{Bensby2013} on the age of 58 microlensed dwarf and subgiant stars in the Galactic bulge by means of completely different stellar models and estimation techniques. 

To this purpose we used the SCEPtER pipeline \citep{scepter1,eta}, which allows  mass, radius, and age of single stars to be estimated by  relying on a fine grid of precomputed stellar tracks and a set of observational constraints. 
Here we adopted the effective temperature, the surface gravity, and [Fe/H] provided by \citet{Bensby2013}. The standard grid was composed by models from ZAMS up to the helium flash at the RGB tip, for mass in the range [0.60; 1.60] $M_{\sun}$. 

The performance of the estimation technique for the considered evolutionary phases was assessed by means of a Monte Carlo simulation. The results of this test showed that the adopted method is unbiased and it can be safely employed for the required age estimations.
  
We confirmed the results by \citet{Bensby2013}, computed adopting a different estimation technique and a different grid of stellar models. This confirmation adds considerable robustness to the age estimates and in particular to the discovery of a sizeable fraction of young bulge stars. We found in fact that the 
analysed sample of microlensed stars contains 16 stars (about one third of the sample) younger than 5 Gyr and 27  (about one half of the sample) younger than 9 Gyr. Even conservatively accounting for the errors in the obtained stellar ages, assuming the upper values as best age estimates, we found that 15 stars were definitely younger than 9 Gyr.

Given the importance of the result and its widespread implications, we analysed in detail possible biases in the age estimates due to some assumptions adopted in computing the grid of stellar models. We first checked if helium rich models would be able to rule out the evidence of the presence of young and 
intermediate age stars as suggested by \citet{Nataf2012}. To this purpose we repeated the age estimation procedure relying on a helium-enhanced grid of models computed on purpose by adopting a helium-to-metal enrichment ratio $\Delta Y/\Delta Z = 5$. The result was a median increase in the age of the stars with [Fe/H]> 0 by about 0.6 Gyr, largely insufficient to reconcile the stellar age estimates with the picture of a totally old bulge suggested by photometric studies of the main-sequence turnoff \citep[e.g.][]{Zoccali2003,Brown2010,Clarkson2011,Valenti2013}. 
To exclude the possibility that the results could arise only by a fluctuation, we also verified the robustness of our finding assuming the 1 $ \sigma $ upper values of age estimates as best values rather then the mean ones. Even with this extreme choice 
and adopting the helium-enhanced grid of stellar models there is evidence of 4 and 15 stars younger than 5 Gyr and 9 Gyr, respectively.

Another  important source of systematic bias in the age estimate of dwarf stars is the treatment of element diffusion \citep{scepter1, eta}. We verified that even neglecting the diffusion in the stellar models adopted in the estimate procedure is not able to increase the age enough to rule out the 
presence of a young and intermediate age population. We obtained a maximum increase of about 1 Gyr for stars younger than 5 Gyr and of about 2.8 Gyr for stars with ages of about 8 Gyr. The mean age increase is indeed much smaller,  about 0.40 Gyr for stars younger than 5 Gyr. 
Even in the most unfavourable case in which the upper values of the age estimates are used, the number of stars younger than 5 and 9 Gyr are, respectively, 4 and 11. 

Finally, we checked the effect of changing the calibration of the mixing-length from $\alpha_{\rm ml} = 1.74$ (our solar value) to $\alpha_{\rm ml} = 1.90$ -- a value calibrated on the RGB colour of old and intermediate age clusters -- in the stellar models used to estimate the age. As in the previous cases, the resulting age increase is not old  enough to reject in a statistically meaningful way the presence of young bulge stars. In fact, we found a maximum age increase of about 1.5 Gyr, and a mean increase of 0.54 Gyr for stars younger than 5 Gyr. 
Even adopting the upper values of the age estimates, the stars younger than 5 and 9 Gyr are, respectively, 4 and 13. 

In conclusion, the presence of a young and intermediate age population among the microlensed bulge appears very robust and statistically confirmed.

\begin{acknowledgements}

We thank the anonymous referee for the valuable comments.
This work has been supported by PRIN-MIUR 2010-2011 ({\em Chemical and dynamical evolution 
of the Milky Way and Local Group galaxies}, PI F. Matteucci) and  PRIN-INAF 2012 
 ({\em The M4 Core Project with Hubble Space Telescope}, PI
L. Bedin). 

\end{acknowledgements}

\bibliographystyle{aa}
\bibliography{biblio}

\appendix
\section{Expected biases on grid estimates}\label{sec:bias}

We 
present here an evaluation of the expected errors and possible bias of the technique adopted for age estimates. 
This information helps to evaluate the accuracy of the results presented in the paper.

For this test we sampled $N = 50\,000$ artificial stars from the recovery grid. To mimic the microlensed sample characteristics, we restricted the sampling to stars up to the very first phase of RGB. The observational parameters of these stars are then subjected to a random Gaussian perturbation, assuming as observational errors 100 K in $T_{\rm eff}$, 0.2 dex in $\log g$, and 0.1 dex in [Fe/H]. The star masses and ages were then estimated by means of the SCEPtER technique. 

\begin{figure*}
        \centering
        \includegraphics[height=17cm,angle=-90]{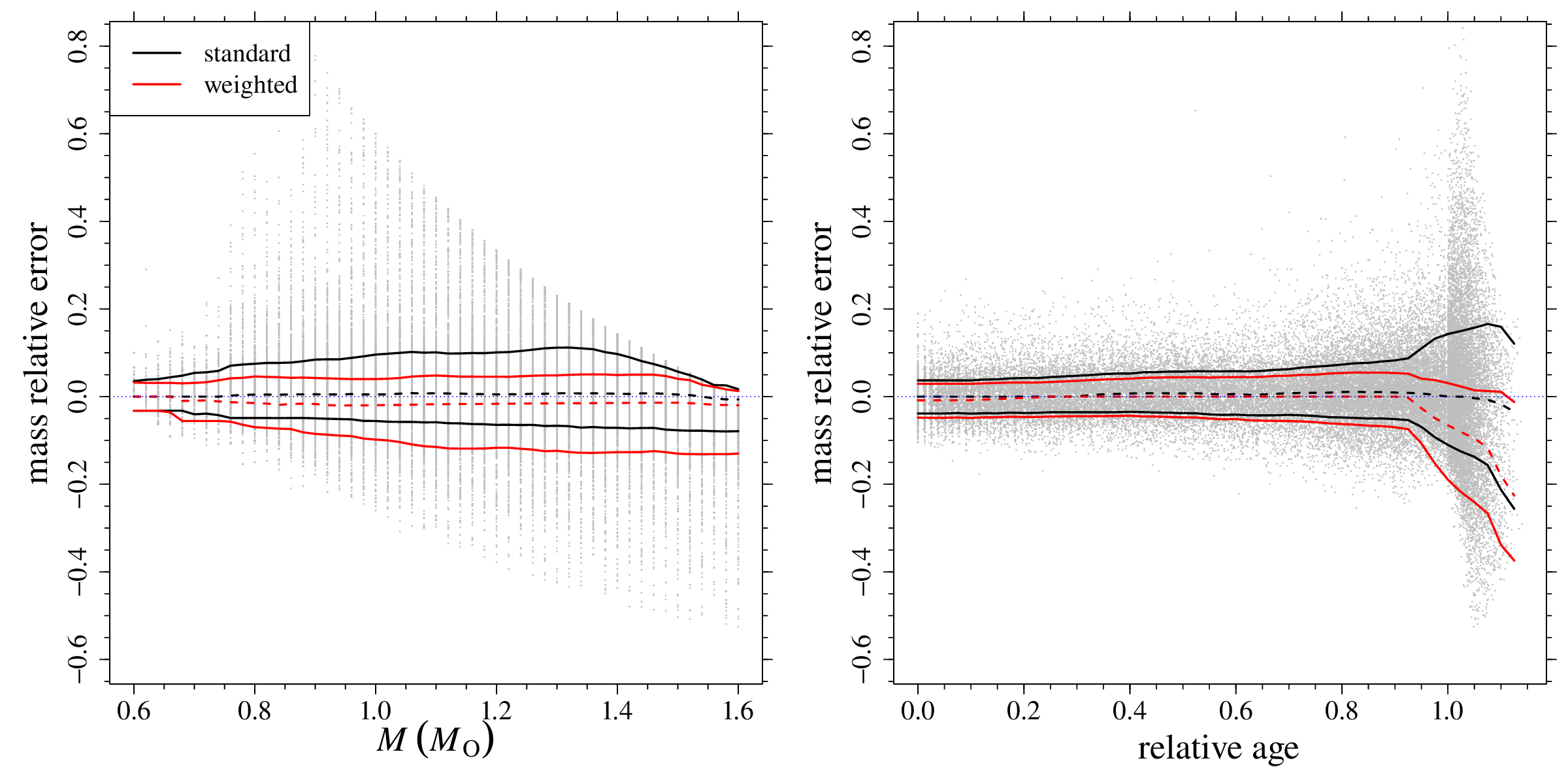}\\
        \includegraphics[height=17cm,angle=-90]{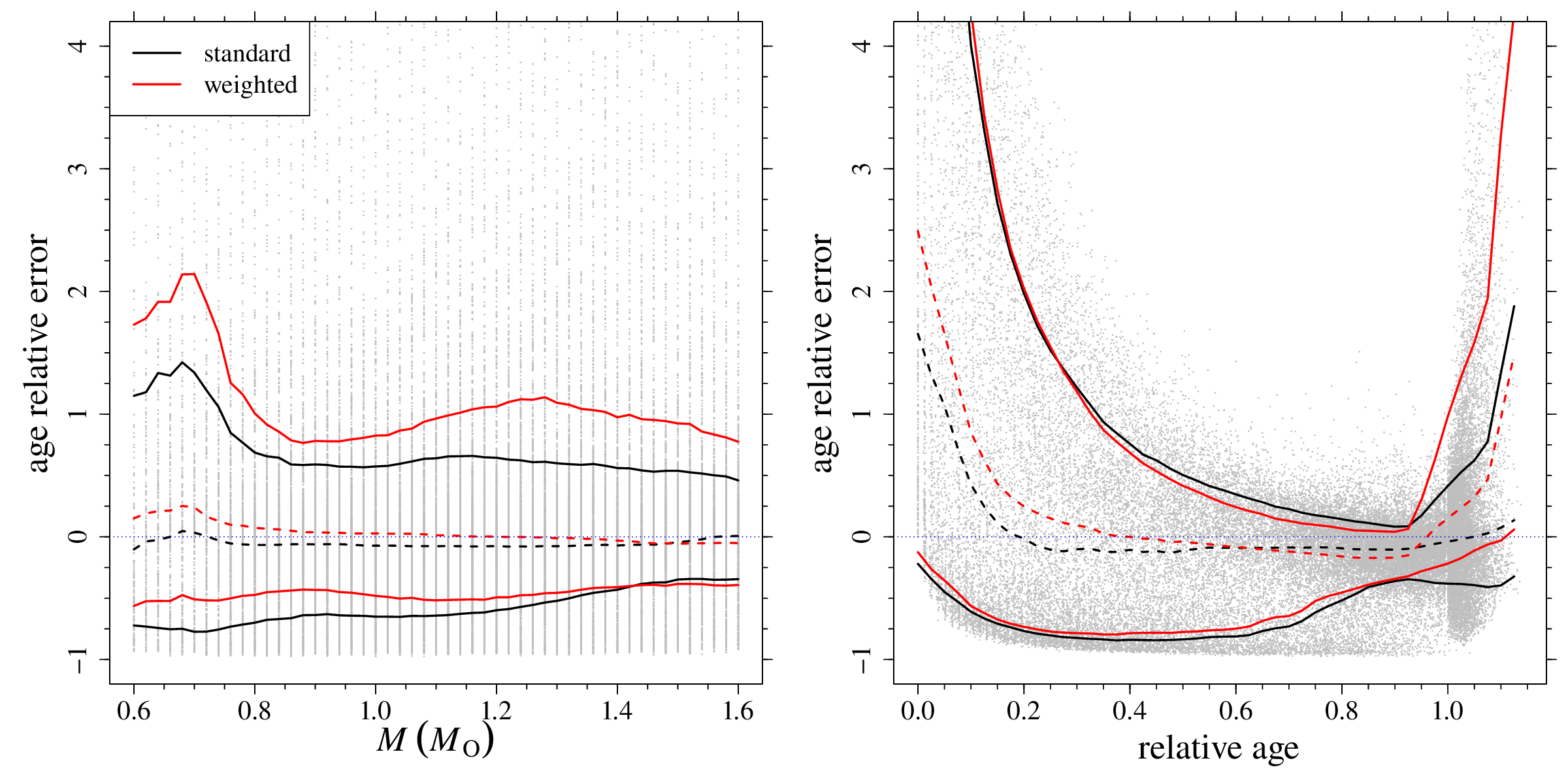}
        \caption{{\it Top row}: {\it left} Monte Carlo relative errors on mass estimate as a function of the true mass of the stars. The black lines show the position of the median (dashed line) and the $1 \sigma$ envelope (solid line). The red lines show the corresponding quantities, but for weighted estimates. {\it Right}: same as in the {\it left panel}, but as a function of the relative age of the stars.  
                {\it Bottom row}: same as the {\it top row}, but for age estimates.} 
        \label{fig:error-mass-age}
\end{figure*}

The resulting relative errors in mass and age are presented in Fig.~\ref{fig:error-mass-age}, where a positive value in the relative error implies that the technique overestimates the considered parameter. 
The figure shows the trend of the relative errors versus the true mass of the star and its relative age\footnote{The relative age is defined as the ratio between
        the current age of the star and its age at 
        the central hydrogen depletion. It is conventionally set to 0 at the ZAMS.}. 
To better follow the resulting trends in the figure we superimpose  the median and the $1 \sigma$ error envelope, computed over a moving window \citep[see][for details on the envelope computation]{eta}.
The figure shows some clear signatures of edge effects, such as the trend in the relative error of mass estimate versus the true mass  (top left panel in Fig.~\ref{fig:error-mass-age}) or the huge distortion of age estimate at low relative age (bottom right panel in the figure), extensively discussed in \citet{scepter1,eta}.
As an example, the trend in the top left panel of the figure at the high mass edge ($M = 1.6$ $M_{\sun}$) originates from the impossibility of obtaining a mass estimate above 1.6 $M_{\sun}$, the maximum value considered in the grid construction. Similar considerations apply to the other grid edges \citep{eta}.  
The tail toward age overestimation, visible at low masses in the bottom left panel, is due to the fact that for stars of mass lower than about 0.7 $M_{\sun}$ the grid does not contain models at high relative age because of the constraint of the 15 Gyr maximum age mentioned in Sect.~\ref{sec:modelgrid}. 
In other words for these masses the grid does not reach  the later central hydrogen burning phases, which leads to more precise age estimates (see bottom right panel in Fig.~\ref{fig:error-mass-age}). The lack of these models causes
the asymmetry in the envelope, typical of the first evolutionary phases.
As expected \citep[see e.g.][]{Gai2011}, mass and age estimates become more difficult in the subgiant phase (relative age higher than 1), since tracks are more closely packed in this grid zone. Moreover, for relative ages higher than about 1.1  a slight mass underestimation distortion appears, with a corresponding increase in age estimate (see the lower envelope boundary in the top right panel of Fig.~\ref{fig:error-mass-age}, and the corresponding upper boundary in the bottom right panel). The trend is due to the packing of the tracks toward the RGB ascension; low mass tracks pack more strictly than those of high mass, therefore the mass is more easily underestimated than overestimated.   

For our purposes, the most interesting feature is the apparent lack of bias in the median of mass and age estimates visible in Fig.~\ref{fig:error-mass-age} (with the exception of the first 20\% of the evolution where edge effects dominate). It also appears from the bottom right panel of the figure that in the subgiant phase the age estimates present a  tail toward overestimation.

It is often stated in the literature that mass and age estimates computed neglecting the evolutionary
time step suffer from a bias and they could be improved by accounting for this correction
\citep[see e.g.][]{Jorgensen2005, Casagrande2011}.
This  statement also appears in \citet{Bensby2013} in relation to the microlensed star age estimates. However, this is not always true since the effect is strictly related to the employed technique and the considered evolutionary stages.
In particular, for the results presented in this paper, the weighted estimates obtained adopting  the evolutionary time step as weight are clearly biased toward age overestimation (bottom right panel in Fig.~\ref{fig:error-mass-age}). The effect is most evident in the subgiant phase, where the bias is higher than 100\%. In this phase, where tracks are closely crowded, the time step weight can make a large difference, leading to the selection of low-mass high-age models.  
Figure~\ref{fig:error-mass-age} shows these trends by reporting the $1 \sigma$ envelope 
of the weighted estimates. Therefore, in the light of this result, in the present work we adopt unweighted estimates as a reference scenario.

\end{document}